\documentclass[conference]{IEEEtran}
\IEEEoverridecommandlockouts

\usepackage{cite}
\usepackage{amsmath,amssymb,amsfonts}
\usepackage{algorithmic}
\usepackage{algorithm}
\usepackage{graphicx}
\usepackage{textcomp}
\usepackage{xcolor}
\usepackage{booktabs}
\usepackage{multirow}
\usepackage{pifont}
\usepackage{tikz}
\usetikzlibrary{arrows.meta, positioning, shapes.geometric, fit, backgrounds, calc, shadows}

\newcommand{\sk}[1]{\text{\textsc{#1}}}

\def\BibTeX{{\rm B\kern-.05em{\sc i\kern-.025em b}\kern-.08em
    T\kern-.1667em\lower.7ex\hbox{E}\kern-.125emX}}

\begin{document}

\title{SkillCom: Decomposing LLM-based Semantic Communication into Task and Channel Aware Skills}

\author{Jingwen Fu, Ming Xiao, Mikael Skoglund\\
KTH Royal Institute of Technology \\
E-mail:~\{jingwenf, mingx, skoglund\}@kth.se
}

\maketitle

\begin{abstract}
Large language models (LLMs) are increasingly used as semantic encoders and decoders in semantic communication. However, current LLM based systems mostly remain monolithic: a single prompted model, or a tightly coupled transmitter/receiver pair, must jointly perform semantic encoding, channel adaptation, and semantic decoding.
Such coupling makes intermediate decisions difficult to control, diagnose, or replace, and may cause channel corruption to propagate through a compressed source representation.
To address the limitations, we propose \textbf{SkillCom}, a modular framework that decomposes LLM-based semantic communication into four explicit skills: semantic abstraction skill, channel-adaptive transmission skill, receiver-side repair skill, and task execution skill.
These skills are interconnected through typed semantic-unit interfaces. Thus, transmission operates on structured unit-level representations rather than on one monolithic text block.
This design localizes channel impairment, enables targeted repair from successfully received units, and supports stage-wise ablation and single-skill replacement under matched communication constraints.
Experiments on multi-hop question answering and dialogue state tracking show that SkillCom consistently outperforms the monolithic LLM baseline, remains more robust under varying channel conditions, and exhibits task-dependent preferences over skill realizations.
The results suggest that explicit skill decomposition provides a more robust and diagnosable foundation for LLM-based semantic communication than monolithic methods.
\end{abstract}

\begin{IEEEkeywords}
Semantic communication, large language models, modular decomposition, task-oriented communication, agent skills
\end{IEEEkeywords}

\section{Introduction}
\label{sec:introduction}

Classical communication theory primarily optimizes reliable bit-level transmission and recovery under channel constraints \cite{bib:shannon_weaver}.
In many artificial intelligence (AI)-driven applications, however, the receiver does not require a faithful copy of the source text; it only requires sufficient task-relevant meanings to produce the correct downstream output \cite{bib:bao_deepsc}.
This motivates \emph{semantic communication}, which transmits meaning rather than raw symbols and can provide a better task-performance/communication-cost tradeoff \cite{bib:xie_deep_sc}.
Representative semantic and task-oriented communication systems have demonstrated the promise of this paradigm. However they are typically end-to-end trained and have a monolithic structure, which makes individual processing stages difficult to isolate, replace, and diagnose \cite{bib:shao_task_oriented, bib:fu_cre_toc, bib:fu_multimodal, bib:semcom_survey}.

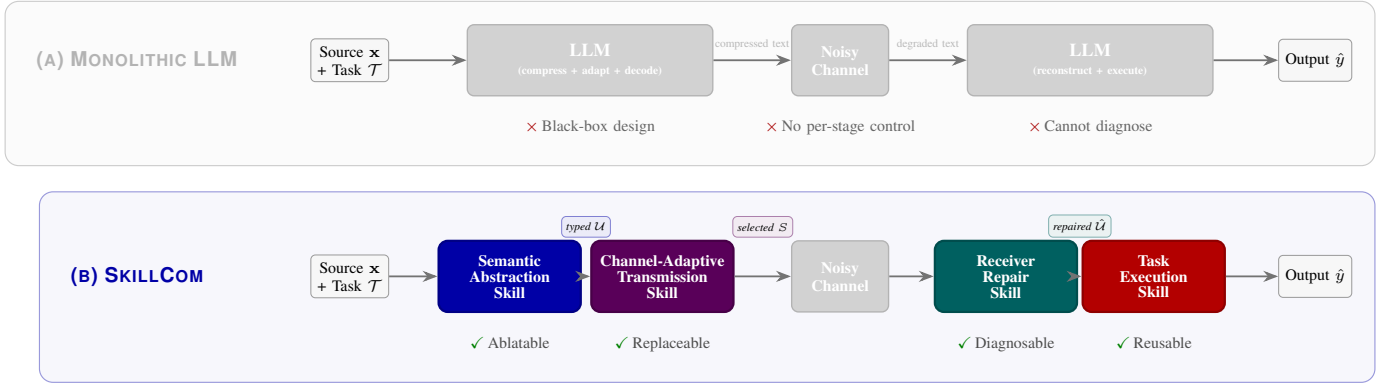
\begin{figure*}[t]
\centering

\colorlet{AbsColor}{blue!65!black}
\colorlet{TransColor}{violet!70!black}
\colorlet{RepairColor}{teal!75!black}
\colorlet{ExecColor}{red!70!black}

\resizebox{\textwidth}{!}{%
\begin{tikzpicture}[
    >=Stealth,
    monoblock/.style={
        rectangle, rounded corners=3pt, draw=gray!60, line width=0.6pt,
        fill=gray!35, minimum height=1.1cm, minimum width=3.8cm,
        font=\footnotesize\bfseries, align=center, text=white
    },
    skillmod/.style={
        rectangle, rounded corners=4pt, draw=#1!80!black, line width=1pt,
        fill=#1, minimum height=1.1cm, minimum width=2.2cm,
        font=\scriptsize\bfseries, align=center, text=white,
        drop shadow={shadow xshift=0.4pt, shadow yshift=-0.4pt, opacity=0.2}
    },
    iobox/.style={
        rectangle, rounded corners=2pt, draw=black!40, line width=0.4pt,
        fill=gray!6, minimum height=0.65cm,
        font=\scriptsize, align=center, inner sep=3pt
    },
    iface/.style={
        rectangle, rounded corners=2pt, draw=#1!50, line width=0.4pt,
        fill=#1!8, minimum height=0.32cm,
        font=\tiny\itshape, align=center, inner sep=2pt
    },
    iface/.default={blue},
    chanmod/.style={
        rectangle, rounded corners=3pt, draw=gray!60, line width=0.6pt,
        fill=gray!35, minimum height=1.1cm, minimum width=1.5cm,
        font=\scriptsize\bfseries, align=center, text=white
    },
    mainarr/.style={->, thick, black!55},
    annot/.style={
        font=\scriptsize, align=center
    },
    rowlabel/.style={
        font=\small\bfseries\scshape, text=#1, align=center
    },
    bgzone/.style={
        rounded corners=6pt, draw=#1!40, line width=0.5pt,
        fill=#1!3, inner sep=10pt
    }
]

\node[rowlabel=gray!60] (label_a) {\textsf{(a) Monolithic LLM}};

\node[iobox, right=1.0cm of label_a] (m_in) {Source $\mathbf{x}$\\+ Task $\mathcal{T}$};

\node[monoblock, right=1.2cm of m_in] (m_llm) {LLM\\[-1pt]{\tiny (compress + adapt + decode)}};

\node[chanmod, right=1.2cm of m_llm] (m_ch) {Noisy\\Channel};

\node[monoblock, right=1.2cm of m_ch] (m_rx) {LLM\\[-1pt]{\tiny (reconstruct + execute)}};

\node[iobox, right=1.0cm of m_rx] (m_out) {Output $\hat{y}$};

\draw[mainarr] (m_in) -- (m_llm);
\draw[mainarr] (m_llm) -- node[above, font=\tiny, text=gray!55, yshift=1pt] {compressed text} (m_ch);
\draw[mainarr] (m_ch) -- node[above, font=\tiny, text=gray!55, yshift=1pt] {degraded text} (m_rx);
\draw[mainarr] (m_rx) -- (m_out);

\node[annot, below=0.25cm of m_llm] (x1) {%
    \color{red!65!black}$\times$\,\color{black!60}Black-box design};
\node[annot, below=0.25cm of m_ch] (x3) {%
    \color{red!65!black}$\times$\,\color{black!60}No per-stage control};
\node[annot, below=0.25cm of m_rx] (x2) {%
    \color{red!65!black}$\times$\,\color{black!60}Cannot diagnose};

\begin{scope}[on background layer]
    \node[bgzone=gray, fit=(label_a)(m_in)(m_llm)(m_ch)(m_rx)(m_out)(x1)(x2)(x3)] (bg_a) {};
\end{scope}

\node[rowlabel=AbsColor, below=2.8cm of label_a] (label_b) {\textsf{(b) SkillCom}};

\node[iobox]   (s_in)  at (m_in  |- label_b) {Source $\mathbf{x}$\\+ Task $\mathcal{T}$};
\node[chanmod] (s_ch)  at (m_ch  |- label_b) {Noisy\\Channel};
\node[iobox]   (s_out) at (m_out |- label_b) {Output $\hat{y}$};

\coordinate (px_abs)   at ($(s_in.east)!0.30!(s_ch.west)$);
\coordinate (px_trans) at ($(s_in.east)!0.68!(s_ch.west)$);
\coordinate (px_rep)   at ($(s_ch.east)!0.30!(s_out.west)$);
\coordinate (px_exec)  at ($(s_ch.east)!0.68!(s_out.west)$);

\node[skillmod={AbsColor}]   (s_abs)   at (px_abs   |- label_b) {Semantic\\[-1pt]Abstraction\\[-1pt]Skill};
\node[skillmod={TransColor}] (s_trans) at (px_trans |- label_b) {Channel-Adaptive\\[-1pt]Transmission\\[-1pt]Skill};
\node[skillmod={RepairColor}]   (s_rep)   at (px_rep   |- label_b) {Receiver\\[-1pt]Repair\\[-1pt]Skill};
\node[skillmod={ExecColor}] (s_exec)  at (px_exec  |- label_b) {Task\\[-1pt]Execution\\[-1pt]Skill};

\path (s_abs.east) -- (s_trans.west) node[midway, above=0.6cm, iface=AbsColor]   (if1) {typed $\mathcal{U}$};
\path (s_trans.east) -- (s_ch.west)  node[midway, above=0.6cm, iface=TransColor] (if2) {selected $S$};
\path (s_rep.east) -- (s_exec.west)  node[midway, above=0.6cm, iface=RepairColor]   (if3) {repaired $\hat{\mathcal{U}}$};

\draw[mainarr] (s_in) -- (s_abs);
\draw[mainarr] (s_abs) -- (s_trans);
\draw[mainarr] (s_trans) -- (s_ch);
\draw[mainarr] (s_ch) -- (s_rep);
\draw[mainarr] (s_rep) -- (s_exec);
\draw[mainarr] (s_exec) -- (s_out);

\node[annot, below=0.25cm of s_abs] (c1) {%
    \color{green!50!black}$\checkmark$\,\color{black!70}Ablatable};
\node[annot, below=0.25cm of s_trans] (c2) {%
    \color{green!50!black}$\checkmark$\,\color{black!70}Replaceable};
\node[annot, below=0.25cm of s_rep] (c3) {%
    \color{green!50!black}$\checkmark$\,\color{black!70}Diagnosable};
\node[annot, below=0.25cm of s_exec] (c4) {%
    \color{green!50!black}$\checkmark$\,\color{black!70}Reusable};

\begin{scope}[on background layer]
    \node[bgzone=AbsColor, fit=(label_b)(s_in)(s_abs)(s_trans)(s_ch)(s_rep)(s_exec)(s_out)(c1)(c2)(c3)(c4)(if1)(if2)(if3)] (bg_b) {};
\end{scope}

\end{tikzpicture}%
}
\caption{Comparison between monolithic semantic communication and SkillCom. The monolithic paradigm treats compression, channel adaptation, and task execution as a single black-box process, whereas SkillCom decomposes the pipeline into four typed skills with explicit interfaces.}
\label{fig:comparison}
\end{figure*}

Large language models (LLMs) have recently emerged as powerful semantic encoders and decoders because of their abilities in abstraction, compression, reasoning, and context regeneration \cite{bib:llm_telecom_survey, Pokhrel2024FLLM}.
Most current LLM-based semantic communication systems, however, remain monolithic, where a single prompted model or a tightly coupled transmitter/receiver LLM pair jointly handles the stages of semantic abstraction, channel encoding, and downstream decoding without explicit stage separation (Fig.~\ref{fig:comparison}(a)) \cite{Pokhrel2024FLLM, bib:wang_wcnc_llm_e2e, bib:lamosc}.
The process creates three limitations.
First, the model must simultaneously handle source compression, channel encoding, and task fidelity within a single prompt, creating conflicting optimization objectives.
Second, the transmitted representation is usually a single text block. Thus, localized channel impairment can disrupt global semantic coherence, leading to catastrophic failure.
Third, because intermediate processing decisions remain black box inside the prompt, failures cannot be attributed to specific stages. Thus, the systems are difficult to diagnose and improve.

These limitations are not caused by the capability of LLMs themselves, but by their monolithic deployment.
Recent work on LLM agents likewise argues that complex behaviors should be organized as explicit and reusable modules at inference time, rather than being left implicit inside a single LLM call \cite{bib:large_model_agents, bib:skillsbench}.
The same principle applies naturally to semantic communication, where semantic abstraction, channel-aware transmission, receiver-side recovery, and task execution are distinct operations with different roles.

Motivated by this observation, we propose \emph{SkillCom}.
To the best of our knowledge, SkillCom is the first framework that decomposes LLM-based semantic communication into four explicit and independently replaceable skills with typed interfaces: semantic abstraction skill, channel-adaptive transmission skill, receiver-side repair skill, and task execution skill.
A central design principle of SkillCom is \emph{independent semantic-unit transmission}: the source is first abstracted into structured semantic units, and the selected units are transmitted independently rather than as one text block.
This converts channel impairment from whole-message failure into localized unit-level erasures, thereby enabling targeted receiver-side repair, stage-wise diagnosis, and modular replacement while the rest of the processing chain remains fixed (Fig.~\ref{fig:comparison}(b)).

The main contributions of this work are summarized as follows:
\begin{enumerate}
    \item We propose SkillCom, to the best of our knowledge the first framework that decomposes LLM-based semantic communication into four explicit and independently replaceable skills, namely semantic abstraction skill, channel-adaptive transmission skill, receiver-side repair skill, and task execution skill, connected through typed interfaces. This formulation enables stage-wise ablation, single-skill replacement, and targeted diagnosis that are difficult to achieve in monolithic semantic systems.
    \item We introduce independent semantic-unit transmission, in which typed semantic units are transmitted individually rather than as a single compressed text block. This localizes channel impairment at the unit level so that packet loss affecting one unit does not invalidate the others, enabling receiver-side repair from successfully received units.
    \item We conduct evaluations on multi-hop question answering (QA) and dialogue state tracking (DST) tasks under channel constraints and varying noise regimes. The results show that SkillCom outperforms monolithic LLM models and that the preferred skill realization depends on different tasks.
\end{enumerate}

\section{System Model and Semantic Representation}
\label{sec:system_model}

\subsection{Problem Formulation}
\label{subsec:problem_formulation}

We consider an end-to-end semantic communication system in which the transmitter observes source text $\mathbf{x}$ and a task descriptor $\mathcal{T}$, and sends task-relevant meaning over a noisy channel to a receiver that produces the output $\hat{y}$.
Transmission is constrained by a communication budget $B$, and the channel state is denoted by $\mathbf{c}$, which we assume is available to both the transmitter and the receiver.
Task performance is measured by a utility function $m(\hat{y}, y^\star)$, where $y^\star$ is the reference output.
The design goal is to maximize task utility under the communication budget:
\begin{equation}
    \max \ \mathbb{E}\!\left[m(\hat{y}, y^\star)\right]
    \qquad
    \text{s.t.}
    \qquad
    \mathrm{cost}(\mathcal{U}_S) \leq B,
\end{equation}
where $\mathcal{U}_S = \{u_j\}_{j \in S}$ denotes the transmitted semantic representation and $\mathrm{cost}(\cdot)$ is its communication cost.

Within this formulation, SkillCom decomposes the transmitter--receiver processing chain into four sequential skills (Fig.~\ref{fig:comparison}(b)).
The transmitter first abstracts and then selects:
\begin{equation}
    \mathcal{U} = \sk{Abs}(\mathbf{x}, \mathcal{T}),
    \label{eq:abstraction}
\end{equation}
where $\mathcal{U}=\{u_1,\ldots,u_N\}$ is the semantic-unit set, and
\begin{equation}
    S = \sk{Trans}(\mathcal{U}, \mathbf{c}, B, \mathcal{T}),
    \label{eq:transmission}
\end{equation}
where $S \subseteq \{1,\ldots,N\}$ is the subset selected for transmission.
The transmitted semantic representation $\mathcal{U}_S = \{u_j\}_{j \in S}$ is sent through the channel, yielding the corrupted received representation $\tilde{\mathcal{U}}_S$.
The receiver then repairs and executes:
\begin{equation}
    \hat{\mathcal{U}} = \sk{Repair}(\tilde{\mathcal{U}}_S, \mathcal{T}),
    \label{eq:repair}
\end{equation}
\begin{equation}
    \hat{y} = \sk{Exec}(\hat{\mathcal{U}}, \mathcal{T}),
    \label{eq:execution}
\end{equation}
where $\hat{\mathcal{U}}$ is the repaired semantic context and $\hat{y}$ is the final task output.

\begin{figure*}[t]
\centering
\resizebox{\textwidth}{!}{%
\begin{tikzpicture}[
    >=Stealth,
    skillhdr/.style={
        rectangle, rounded corners=4pt, draw=#1!80!black, line width=1pt,
        fill=#1, minimum height=0.55cm,
        font=\footnotesize\bfseries, align=center, text=white,
        drop shadow={shadow xshift=0.3pt, shadow yshift=-0.3pt, opacity=0.15}
    },
    step/.style={
        rectangle, rounded corners=3pt, draw=#1!65!black, line width=0.5pt,
        fill=#1!10, minimum height=0.55cm, minimum width=1.1cm,
        font=\scriptsize, align=center, inner sep=4pt
    },
    step/.default={blue},
    altmode/.style={
        rectangle, rounded corners=2pt, draw=#1!55, line width=0.4pt,
        fill=#1!6, minimum height=0.42cm,
        font=\tiny, align=center, inner sep=2.5pt
    },
    altmode/.default={blue},
    iface/.style={
        rectangle, rounded corners=2pt, draw=#1!50, line width=0.4pt,
        fill=#1!8, minimum height=0.35cm,
        font=\tiny\itshape, align=center, inner sep=2pt
    },
    iface/.default={gray},
    io/.style={
        rectangle, rounded corners=2pt, draw=black!40, line width=0.4pt,
        fill=gray!6, minimum height=0.5cm,
        font=\scriptsize, align=center, inner sep=3pt
    },
    chanmod/.style={
        rectangle, rounded corners=4pt, draw=gray!60, line width=0.7pt,
        fill=gray!35, minimum height=0.9cm, minimum width=1.8cm,
        font=\scriptsize\bfseries, align=center, text=white
    },
    mathlbl/.style={font=\tiny, text=black!70, align=center},
    mainarr/.style={->, thick, black!50},
    dasharr/.style={->, thin, dashed, #1!65!black},
    dasharr/.default={orange},
    panelbg/.style={
        rounded corners=6pt, draw=#1!40, line width=0.5pt,
        fill=#1!3, inner sep=0pt
    }
]


\begin{scope}[on background layer]
    \node[panelbg=blue,   minimum width=6.5cm, minimum height=3.4cm] at ( 1.75, 0)    {};
    \node[panelbg=violet, minimum width=6.5cm, minimum height=3.4cm] at ( 9.75, 0)    {};
    \node[panelbg=teal,   minimum width=6.5cm, minimum height=3.0cm] at ( 9.75,-4.5)  {};
    \node[panelbg=red,    minimum width=6.5cm, minimum height=3.0cm] at ( 1.75,-4.5)  {};
\end{scope}

\node[skillhdr=blue!65!black] at (1.75, 1.3) {Semantic Abstraction};

\node[io] (src) at (-0.3, 0) {Source $\mathbf{x}$,\;\!Task $\mathcal{T}$};
\node[step=blue] (abs) at (2.0, 0) {Task-Aware\\[-1pt]Extraction};

\node[iface=blue] (ut1) at (4.0,  0.48) {Keyword};
\node[iface=blue] (ut2) at (4.0,  0.16) {Entity};
\node[iface=blue] (ut3) at (4.0, -0.16) {Evidence};
\node[iface=blue] (ut4) at (4.0, -0.48) {Slot-Val};

\node[altmode=blue] (m1) at (0.6, -1.15) {\emph{(i)} Heuristic\\[-1pt](0 LLM)};
\node[altmode=blue] (m2) at (1.9, -1.15) {\emph{(ii)} LLM-\\[-1pt]Enriched};
\node[altmode=blue] (m3) at (3.2, -1.15) {\emph{(iii)} Structured\\[-1pt]LLM};

\draw[mainarr] (src) -- (abs);
\foreach \u in {ut1,ut2,ut3,ut4}
    \draw[->,thin,blue!50] (abs.east) -- ++(0.12,0) |- (\u.west);
\draw[dasharr=blue] (m1.north) -- (abs.south);
\draw[dasharr=blue] (m2.north) -- (abs.south);
\draw[dasharr=blue] (m3.north) -- (abs.south);

\node[io] (ifU) at (5.75, 0)
    {$\mathcal{U}{=}\{u_j\}$\\[-1pt]{\tiny$(\xi,\tau,r,s,g,\kappa)$}};
\draw[mainarr] (ut1.east) -- ++(0.1,0) |- (ifU.west);
\draw[mainarr] (ut4.east) -- ++(0.1,0) |- (ifU.west);

\node[skillhdr=violet!70!black] at (9.75, 1.3) {Channel-Adaptive Transmission};

\node[step=violet] (prio)  at ( 7.6, 0) {Priority\\[-1pt]Scoring};
\node[step=violet] (gsel)  at ( 9.0, 0) {Greedy\\[-1pt]Selection};
\node[step=violet] (dedup) at (10.4, 0) {LLM\\[-1pt]Dedup};
\node[step=violet] (budg)  at (11.8, 0) {Budget\\[-1pt]Enforce};

\node[mathlbl] at (9.6, -1.2)
    {$w_j{=}\alpha_r r_j{+}\alpha_s s_j{+}\alpha_g g_j{-}\alpha_c\kappa_j$,};
\node[mathlbl] at (9.6, -1.5)
    {st. $|S|{\le}B_u,\;\Sigma\kappa{\le}B_\kappa,\;\Sigma|\xi|{\le}B_c$};

\node[iface=violet] (snr) at (10.4, -0.8) {SNR};
\draw[dasharr=violet] (snr) -- (dedup);

\draw[mainarr] (ifU) -- (prio);
\draw[mainarr] (prio) -- (gsel);
\draw[mainarr] (gsel) -- (dedup);
\draw[mainarr] (dedup) -- (budg);

\node[io] (ifS) at (13.8, 0) {$S{\subseteq}\{1{\ldots}N\}$};
\draw[mainarr] (budg) -- (ifS);

\node[chanmod] (chan) at (13.8, -2.3)
    {Erasure Channel\\[-2pt]{\tiny$\mathrm{PER}_j{=}1{-}(1{-}p_b)^{n_j}$}};
\draw[mainarr] (ifS) -- (chan);

\node[io] (ifUtilde) at (13.8, -4.5)
    {$\tilde{\mathcal{U}}_S$\\[-1pt]{\tiny successfully received units}};
\draw[mainarr] (chan) -- (ifUtilde);

\node[skillhdr=teal!75!black] at (9.75, -3.4) {Receiver Repair};

\node[step=teal] (edet) at (11.5, -4.5) {Erasure\\[-1pt]Detection};
\node[altmode=teal] (rp1) at (8.8, -4.1)
    {\emph{(i)} Generative:\\[-1pt]synthesize units};
\node[altmode=teal] (rp2) at (8.8, -4.9)
    {\emph{(ii)} Guided:\\[-1pt]context guidance};
\node[mathlbl] at (10, -5.5) {$S_\varnothing{=}S \setminus S_{\mathrm{rx}}$};

\draw[mainarr] (ifUtilde) -- (edet);
\draw[->,thin,teal!50] (edet.west) -- ++(-.12,0) |- (rp1.east);
\draw[->,thin,teal!50] (edet.west) -- ++(-.12,0) |- (rp2.east);

\node[io] (ifUhat) at (5.75, -4.5) {$\hat{\mathcal{U}}$};
\draw[mainarr] (rp1.west) -- ++(-.12,0) |- (ifUhat.east);
\draw[mainarr] (rp2.west) -- ++(-.12,0) |- (ifUhat.east);

\node[skillhdr=red!70!black] at (1.75, -3.4) {Task Execution}; 

\node[step=red!70!black] (exec) at (2.5, -4.5) {LLM\\[-1pt]Decoder};
\node[altmode=red] (tqa)  at (1.5, -5.5) {QA: answer};
\node[altmode=red] (tdst) at (3.5, -5.5) {DST: slots};
\draw[dasharr=red] (tqa.north) -- (exec.south);
\draw[dasharr=red] (tdst.north) -- (exec.south);

\draw[mainarr] (ifUhat) -- (exec);

\node[io] (out) at (-0.3, -4.5) {Output $\hat{y}$};
\draw[mainarr] (exec) -- (out);

\node[font=\footnotesize\scshape, text=black!35, rotate=90, anchor=center]
    at (-2.2,  0)   {Sender};
\node[font=\footnotesize\scshape, text=black!35, rotate=90, anchor=center]
    at (-2.2, -4.5) {Receiver};

\end{tikzpicture}%
}
\caption{One realization of the proposed SkillCom processing chain. The transmitter abstracts the source into typed semantic units and selects units under budget constraints, the channel independently erases transmitted units, and the receiver repairs the received units before task execution.}
\label{fig:skill_details}
\end{figure*}
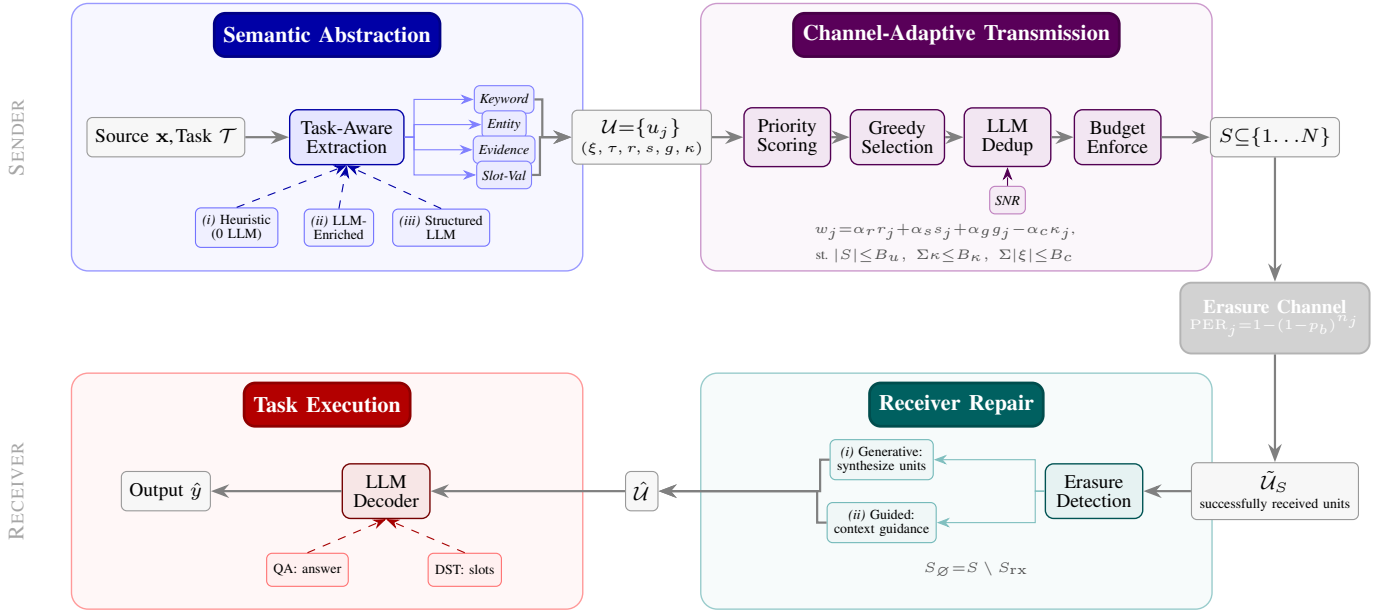

\subsection{Channel Model}
\label{subsec:channel_model}

The channel state $\mathbf{c}$ introduced in Section~\ref{subsec:problem_formulation} is instantiated as the operating signal-to-noise ratio (SNR).
Following the coded packet erasure model over additive white Gaussian noise (AWGN)~\cite{bib:proakis}, each selected semantic unit is treated as an independently coded packet protected by an error-detecting code (e.g., CRC).
For binary phase-shift keying (BPSK) modulation, the bit error rate is given by
\begin{equation}
    p_b = \tfrac{1}{2}\,\mathrm{erfc}\!\bigl(\!\sqrt{10^{\mathrm{SNR}/10}}\bigr).
    \label{eq:ber}
\end{equation}
For a unit $u_j$, let $L_j$ denote its byte length and $g_j$ denote its robustness score.
Its coded length is then
\begin{equation}
    n_j = \bigl\lceil L_j \cdot 8 \cdot (2 - g_j)\,/\,R \bigr\rceil
    \label{eq:coded_length}
\end{equation}
bits, where $R$ is the code rate that accounts for the error-detection overhead.
The factor $(2-g_j)$ adjusts the coded length according to unit robustness.
Because the code detects but does not correct bit errors, the packet is declared erased whenever any bit error is detected.
The resulting packet erasure probability (PER) of $u_j$ is therefore 
\begin{equation}
    \mathrm{PER}_j = 1 - (1 - p_b)^{n_j},
    \label{eq:per}
\end{equation}
and the unit is delivered intact otherwise \cite{bib:proakis}.
The received representation therefore factorizes as
\begin{equation}
    p\!\left(\tilde{\mathcal{U}}_S \mid \mathcal{U}_S, \mathrm{SNR}\right)
    = \prod_{j \in S} p\!\left(\tilde{u}_j \mid u_j, \mathrm{SNR}\right),
    \label{eq:channel_factorize}
\end{equation}
where $p(\tilde{u}_j \mid u_j, \mathrm{SNR})$ is the packet-level transition distribution induced by $\mathrm{PER}_j$.
In particular, $p(\tilde{u}_j = \emptyset \mid u_j, \mathrm{SNR}) = \mathrm{PER}_j$ and $p(\tilde{u}_j = u_j \mid u_j, \mathrm{SNR}) = 1 - \mathrm{PER}_j$.

\subsection{Semantic Unit Representation}
\label{subsec:semantic_units}

In a monolithic system, the transmitter sends a single compressed text block.
This design has a clear limitation: a single packet error can corrupt the entire message.
To address this problem, we propose a \emph{semantic unit representation} for transmission.
Each unit is typed and can be selected, transmitted, and repaired independently.
Formally, one unit is defined as
\begin{equation}
    u_i = \bigl(\xi_i,\; \tau_i,\; r_i,\; s_i,\; g_i,\; \kappa_i \bigr),
    \label{eq:semantic_unit}
\end{equation}
where $\xi_i$ is the semantic payload, i.e., the text content carried by the unit, $\tau_i \in \mathcal{V}$ is the unit type, $r_i \in [0,1]$ is task relevance, $s_i \in [0,1]$ is source importance, $g_i \in [0,1]$ is channel robustness, and $\kappa_i > 0$ is the token cost.
In the transmission stage, ($r_j$, $s_j$, $g_j$, $\kappa_j$) will be used to form the priority score in~\eqref{eq:priority}.
Although the type set $\mathcal{V}$ may vary across tasks, the six-field interface is fixed and shared by all downstream skills.
This fixed interface enables composability across tasks and implementations.

This representation provides two advantages.
First, each unit carries both semantic value and transmission cost, so the transmission skill can perform selection under explicit rate constraints.
Second, this representation supports a unit-level transmission model rather than a single monolithic text block.
As a result, erasure of one unit does not invalidate the remaining received units, which can still support downstream repair and task execution.

\section{Skill Modules and Realizations}
\label{sec:skill_design}

Building on the system design in Section~\ref{sec:system_model}, this section describes representative realizations of the four skills in this paper.
Fig.~\ref{fig:skill_details} shows one realization of these skills.

\subsection{Semantic Abstraction Skill}
\label{subsec:abstraction}

The abstraction skill $\sk{Abs}(\mathbf{x}, \mathcal{T}) \rightarrow \mathcal{U}$ maps raw source text to typed semantic units in~\eqref{eq:semantic_unit}.
Because SkillCom is task-aware, the abstraction stage should adapt to different task needs, such as precise evidence extraction or broader contextual coverage.
We therefore design three realizations of the abstraction skill.
Although they differ in LLM involvement, all produce the same six-field interface $(\xi_i, \tau_i, r_i, s_i, g_i, \kappa_i)$ and are interchangeable for downstream skills.

\paragraph{Heuristic Abstraction}
This realization relies on deterministic extraction heuristics.
Candidate spans are scored using term frequency-inverse document frequency (TF-IDF) salience and entity cues and are then mapped to task-appropriate unit types.
For QA, it emphasizes evidence-bearing spans; for DST, it emphasizes domain--slot--value structure.

\paragraph{LLM-Enriched Abstraction}
This realization augments heuristic extraction with LLM-generated keywords, entities, and concise summaries.
It broadens coverage for implicit or paraphrased information while preserving the same typed interface for downstream skills.

\paragraph{Structured LLM Extraction}
This realization performs abstraction through a JSON-schema-constrained LLM call that directly outputs the target unit types, providing an LLM-driven abstraction process while preserving the shared semantic-unit interface.

As we will show in Section~\ref{sec:experiments}, different tasks favor different abstraction modes, further motivating the task-aware design of the abstraction skill.

\subsection{Channel-Adaptive Transmission Skill}
\label{subsec:transmission}

The transmission skill $\sk{Trans}(\mathcal{U}, \mathbf{c}, B, \mathcal{T}) \rightarrow S$ selects a subset of semantic units for transmission under channel and budget constraints.
In the channel model of Section~\ref{subsec:channel_model}, the channel state $\mathbf{c}$ is instantiated by $\mathrm{SNR}$.
Because different units contribute differently to task success, communication efficiency, and robustness, the selector should prioritize information that is both semantically useful and economical to transmit.
We therefore formulate transmission as the following constrained utility maximization problem:
\begin{equation}
\begin{aligned}
    \max_{S \subseteq \{1,\ldots,N\}} \quad & \sum_{j \in S} w_j \\
    \text{s.t.} \quad & |S| \leq B_u,\;
    \sum_{j \in S} \kappa_j \leq B_\kappa,\;
    \sum_{j \in S} |\xi_j| \leq B_c
\end{aligned}
\label{eq:selection}
\end{equation}
where $|\xi_j|$ denotes the character length of unit $u_j$.
The unit priority score is defined as
\begin{equation}
    w_j = \alpha_r \, r_j + \alpha_s \, s_j + \alpha_g \, g_j - \alpha_c \, \kappa_j,
    \label{eq:priority}
\end{equation}
with non-negative weights $(\alpha_r, \alpha_s, \alpha_g, \alpha_c)$ for task relevance $r_j$, source importance $s_j$, channel robustness $g_j$, and transmission cost $\kappa_j$.
This score favors units that are task-relevant, source-important, and robust to channel impairment, while penalizing transmission cost.
We consider two realizations of the transmission skill.

\paragraph{Greedy Selection}
Units are ranked by $w_j$ in descending order and added to $S$ until adding another unit would exceed one of the budget constraints in~\eqref{eq:selection}.

\paragraph{Greedy Selection with LLM Deduplication}
This realization augments greedy selection with a semantic deduplication and reranking step to reduce redundancy before final budget enforcement.
The resulting selected set still satisfies the constraints in~\eqref{eq:selection}.

\subsection{Receiver Repair Skill}
\label{subsec:repair}

The repair skill $\sk{Repair}(\tilde{\mathcal{U}}_S, \mathcal{T}) \rightarrow \hat{\mathcal{U}}$ reconstructs a semantically usable representation from the successfully received units.
Let $S_{\mathrm{rx}} \subseteq S$ denote the indices of successfully received units and $S_{\varnothing} = S \setminus S_{\mathrm{rx}}$ the erased ones.
From the received unit IDs, the repair skill identifies the missing set $S_{\varnothing}$ and uses the successfully received units $\{u_j\}_{j \in S_{\mathrm{rx}}}$ together with the task descriptor $\mathcal{T}$ to recover missing information or provide structured support for downstream inference.

We consider two task-dependent realizations of the repair skill.

\paragraph{Generative Repair}
This strategy synthesizes replacement units through an LLM conditioned on the successfully received units and the task query.
It is suited to tasks in which partial recovery of missing evidence can still improve downstream inference.

\paragraph{Guided Repair}
This strategy does not synthesize new units.
Instead, it derives structured guidance from the surviving context, such as active domains and confirmed slot--value pairs, and passes this guidance to the execution stage.
It is better suited to tasks that require conservative structured outputs and are sensitive to hallucinated content.

In both cases, the repaired output $\hat{\mathcal{U}}$ remains in the semantic-unit space, so communication recovery and task inference remain separately diagnosable.

\subsection{Task Execution Skill}
\label{subsec:execution}

The execution skill $\sk{Exec}(\hat{\mathcal{U}}, \mathcal{T}) \rightarrow \hat{y}$ maps the repaired semantic representation to the final task output through a task-conditioned LLM decoder.
It is the final stage of the pipeline and the only stage that commits to the downstream output space.
For the QA task, the decoder takes the repaired units together with the query and generates a short answer.
For the DST task, it takes the repaired dialogue representation and outputs canonical slot--value pairs.
Because the execution skill always takes the same repaired-unit interface as input, changes in abstraction, transmission, or repair do not require changing the execution stage.

\section{Numerical Results}
\label{sec:experiments}

\subsection{Experimental Setup}
\label{subsec:experimental_setups}

\subsubsection{Tasks}
We evaluate on HotpotQA \cite{bib:hotpotqa}, a multi-hop question answering benchmark, and MultiWOZ~2.4 \cite{bib:multiwoz}, a multi-domain DST benchmark.

\subsubsection{Methods Under Comparison}
We compare SkillCom with a monolithic baseline.
The monolithic baseline compresses the full source into a single text block at the transmitter and performs task decoding at the receiver without explicit stage decomposition.
We also instantiate SkillCom in four variants to examine which skill configurations are more effective for different tasks.
SkillCom-Heuristic uses heuristic abstraction only, SkillCom-Enrich augments heuristic abstraction with LLM-generated keywords, entities, and summaries, SkillCom-Struct uses a structured LLM call to extract semantic units directly, and SkillCom-Struct+Dedup further adds channel-aware LLM deduplication before transmission.

\subsubsection{Channel and Budget}
All comparisons use the coded packet erasure channel in Section~\ref{subsec:channel_model} with coding rate $R{=}0.5$ and matched budgets across methods.
HotpotQA uses $(B_u, B_\kappa, B_c) = (4, 48, 300)$, and MultiWOZ uses $(5, 56, 350)$.
\subsubsection{Metrics and Implementation}
For HotpotQA we report exact match (EM) and token-level F1; for MultiWOZ we report joint goal accuracy (JGA) and slot-level F1.
We also report the number of LLM calls per sample and transmitted tokens.
For each task, we evaluate $100$ sampled examples with a fixed seed.
All LLM calls use DeepSeek-Chat with temperature~$0$ and caching for reproducibility.
The repair skill uses the generative strategy for HotpotQA and the guided strategy for MultiWOZ.
\subsection{Main Comparison}
\label{subsec:main_positioning}

\begin{table}[t]
\centering
\caption{Main comparison on HotpotQA ($\mathrm{SNR}{=}7\,\text{dB}$).}
\label{tab:main_qa}
\scriptsize
\setlength{\tabcolsep}{3.2pt}
\begin{tabular}{@{}l c c c c@{}}
\toprule
\textbf{Method} & \textbf{EM}$\uparrow$ & \textbf{F1}$\uparrow$ & \textbf{LLM Calls}$\downarrow$ & \textbf{Tx Tokens}$\downarrow$ \\
\midrule
Monolithic        & 0.42 & 0.51 & 2 & 24.0 \\
\midrule
SkillCom-Heuristic    & 0.42 & 0.52 & 2 & 27.9 \\
SkillCom-Enrich       & 0.45 & 0.57 & 5 & 30.0 \\
SkillCom-Struct       & 0.49 & 0.59 & 3 & 19.4 \\
SkillCom-Struct+Dedup & \textbf{0.56} & \textbf{0.68} & 4 & \textbf{6.8} \\
\bottomrule
\end{tabular}
\end{table}

\begin{table}[t]
\centering
\caption{Main comparison on MultiWOZ DST ($\mathrm{SNR}{=}7\,\text{dB}$).}
\label{tab:main_dst}
\scriptsize
\setlength{\tabcolsep}{3.2pt}
\begin{tabular}{@{}lcccc@{}}
\toprule
\textbf{Method} & \textbf{JGA}$\uparrow$ & \textbf{Slot F1}$\uparrow$ & \textbf{LLM Calls}$\downarrow$ & \textbf{Tx Tokens}$\downarrow$ \\
\midrule
Monolithic        & 0.02 & 0.03 & 2 & 26.8 \\
\midrule
SkillCom-Heuristic    & 0.05 & 0.36 & 2 & 17.5 \\
SkillCom-Enrich       & \textbf{0.08} & \textbf{0.42} & 5 & 15.9 \\
SkillCom-Struct       & 0.07 & 0.33 & 3 & 8.3 \\
SkillCom-Struct+Dedup & 0.01 & 0.25 & 4 & \textbf{5.4} \\
\bottomrule
\end{tabular}
\end{table}

Tables~\ref{tab:main_qa} and~\ref{tab:main_dst} present the main comparison at $\mathrm{SNR}{=}7\,\text{dB}$.
First, SkillCom consistently outperforms the monolithic baseline.
On HotpotQA, SkillCom-Struct+Dedup improves EM from $0.42$ to $0.56$ and F1 from $0.51$ to $0.68$.
On MultiWOZ, the monolithic baseline attains only $0.02$ JGA and $0.03$ Slot~F1, whereas all SkillCom variants recover meaningful dialogue state; SkillCom-Enrich achieves $0.08$ JGA and $0.42$ Slot~F1.
Second, the preferred skill realization is task-dependent: structured abstraction with deduplication performs best on HotpotQA, whereas the LLM-enriched variant performs best on MultiWOZ, suggesting that DST benefits more from broad contextual coverage than from aggressive structural compression.
These results show that skill decomposition enables not only modularity but also task-dependent optimization of individual skill implementations.

\subsection{Noise Robustness}
\label{subsec:regime_dependent_utility}

\begin{figure*}[t]
    \centering
    \includegraphics[width=\textwidth]{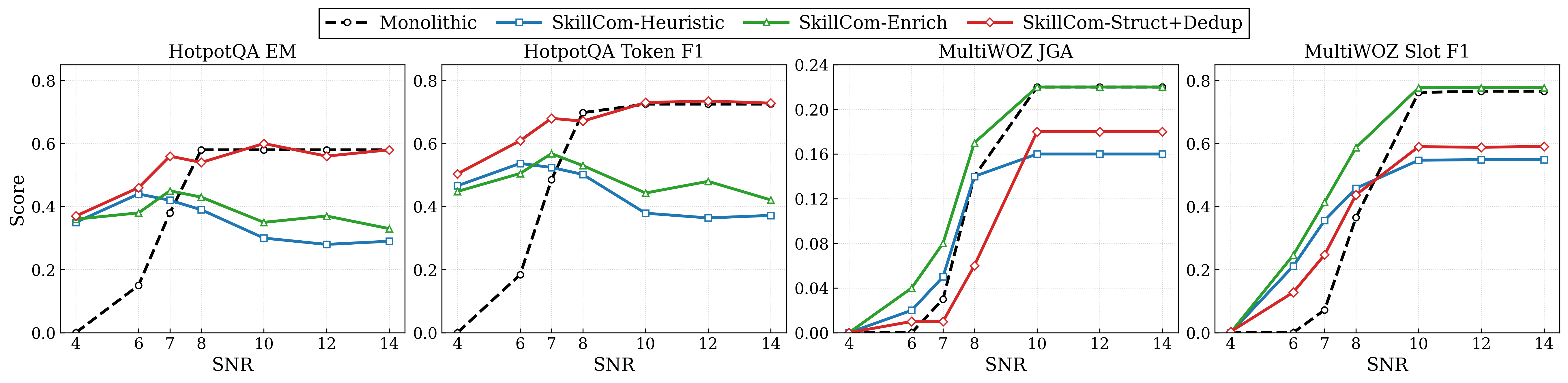}
    \caption{Noise robustness of the monolithic baseline and SkillCom variants across SNR levels on HotpotQA and MultiWOZ.}
    \label{fig:noise_robustness}
\end{figure*}

Fig.~\ref{fig:noise_robustness} summarizes performance over SNR values ranging from $4$ to $14$~dB for the monolithic baseline and three representative SkillCom variants on HotpotQA and MultiWOZ.
First, SkillCom degrades more gracefully than the monolithic baseline as noise increases.
Across both tasks, the monolithic baseline exhibits severe performance degradation as SNR decreases, whereas the SkillCom variants maintain substantially higher utility over the full sweep.
This contrast is especially visible in HotpotQA, where SkillCom-Struct+Dedup remains effective even at low SNR while the monolithic baseline suffers catastrophic degradation near the $\mathrm{SNR}{=}7\text{--}8\,\text{dB}$ regime.
Second, at $\mathrm{SNR}{=}4\,\text{dB}$ the monolithic baseline drops to near-zero performance, while SkillCom still preserves meaningful task performance on both tasks, confirming the structural advantage of unit-level transmission in which erasure of one unit does not invalidate the others.
The task-dependent preference over skill realizations also persists across all noise levels.

\subsection{Skill Ablation}
\label{subsec:ablation}

To quantify the contribution of individual skills, we ablate one component at a time from the best-performing SkillCom variant on each task while keeping all other skills fixed, using the same $\mathrm{SNR}{=}7\,\text{dB}$ setting as the main comparison.
Tables~\ref{tab:ablation_qa} and~\ref{tab:ablation_dst} summarize the results.

\begin{table}[t]
\centering
\caption{Skill ablation on HotpotQA ($\mathrm{SNR}{=}7\,\text{dB}$). Full model: SkillCom-Struct+Dedup.}
\label{tab:ablation_qa}
\scriptsize
\setlength{\tabcolsep}{2.2pt}
\begin{tabular}{@{}ccccccrr@{}}
\toprule
\textbf{LLM Abs.} & \textbf{Dedup} & \textbf{Ch-Aware} & \textbf{Repair} & \textbf{EM}$\uparrow$ & \textbf{F1}$\uparrow$ & $\Delta$\textbf{EM} & $\Delta$\textbf{F1} \\
\midrule
\ding{51} & \ding{51} & \ding{51} & \ding{51} & \textbf{0.56} & \textbf{0.68} & --- & --- \\
\ding{51} & \ding{51} & \ding{51} & \ding{55} & 0.51 & 0.61 & $-0.05$ & $-0.07$ \\
\ding{51} & \ding{51} & \ding{55} & \ding{51} & 0.36 & 0.48 & $-0.20$ & $-0.20$ \\
\ding{51} & \ding{55} & \ding{51} & \ding{51} & 0.49 & 0.59 & $-0.07$ & $-0.09$ \\
\ding{55} & \ding{55} & \ding{51} & \ding{51} & 0.42 & 0.52 & $-0.14$ & $-0.16$ \\
\bottomrule
\end{tabular}
\end{table}

\begin{table}[t]
\centering
\caption{Skill ablation on MultiWOZ DST ($\mathrm{SNR}{=}7\,\text{dB}$). Full model: SkillCom-Enrich.}
\label{tab:ablation_dst}
\scriptsize
\setlength{\tabcolsep}{2.4pt}
\begin{tabular}{@{}ccccc rr@{}}
\toprule
\textbf{LLM Abs.} & \textbf{Ch-Aware} & \textbf{Repair} & \textbf{JGA}$\uparrow$ & \textbf{Slot F1}$\uparrow$ & $\Delta$\textbf{JGA} & $\Delta$\textbf{Slot F1} \\
\midrule
\ding{51} & \ding{51} & \ding{51} & \textbf{0.08} & \textbf{0.42} & --- & --- \\
\ding{55} & \ding{51} & \ding{51} & 0.05 & 0.36 & $-0.03$ & $-0.06$ \\
\ding{51} & \ding{55} & \ding{51} & 0.08 & 0.41 & $\pm 0.00$ & $-0.01$ \\
\ding{51} & \ding{51} & \ding{55} & 0.07 & 0.43 & $-0.01$ & $+0.01$ \\
\bottomrule
\end{tabular}
\end{table}
On HotpotQA (Table~\ref{tab:ablation_qa}), the full model is SkillCom-Struct+Dedup.
Removing channel-aware transmission causes the largest drop ($\Delta\text{F1}{=}{-}0.20$), followed by removing LLM abstraction ($-0.16$) and deduplication ($-0.09$).
This indicates that channel-aware unit prioritization is the most critical skill for QA at this operating point.
On MultiWOZ (Table~\ref{tab:ablation_dst}), the full model is SkillCom-Enrich.
The dominant factor is LLM-enriched abstraction: removing it reduces Slot~F1 from $0.42$ to $0.36$ and JGA from $0.08$ to $0.05$, whereas removing channel-aware transmission or repair has only minor effect.
Overall, the performance-limiting skill varies with the task: channel-aware transmission is most critical for QA, whereas LLM-enriched abstraction is most critical for DST.
This kind of targeted diagnosis is difficult to obtain from a monolithic system and illustrates the analytical value of explicit skill decomposition.

\subsection{Budget Sensitivity}
\label{subsec:budget_sensitivity}

To examine how performance varies with the communication budget, we evaluate the monolithic baseline and two representative SkillCom variants across four budget levels at $\mathrm{SNR}{=}7\,\text{dB}$ (Tables~\ref{tab:budget_qa} and~\ref{tab:budget_dst}).
For each task, the best-performing variant from Section~\ref{subsec:main_positioning} is included alongside the best variant from the other task for cross-comparison.
Budgets are scaled proportionally from a tight setting $(B_u{=}2)$ to a generous setting $(B_u{=}6\text{--}7)$.

\begin{table}[t]
\centering
\caption{Budget sensitivity on HotpotQA ($\mathrm{SNR}{=}7\,\text{dB}$). Budget format: $(B_u, B_\kappa, B_c)$.}
\label{tab:budget_qa}
\scriptsize
\setlength{\tabcolsep}{2.0pt}
\begin{tabular}{@{}l cccc cccc@{}}
\toprule
 & \multicolumn{4}{c}{\textbf{EM}$\uparrow$} & \multicolumn{4}{c}{\textbf{Token F1}$\uparrow$} \\
\cmidrule(lr){2-5} \cmidrule(lr){6-9}
\textbf{Method} & \scriptsize\textbf{Tight} & \scriptsize\textbf{Med.} & \scriptsize\textbf{Def.} & \scriptsize\textbf{Gen.} & \scriptsize\textbf{Tight} & \scriptsize\textbf{Med.} & \scriptsize\textbf{Def.} & \scriptsize\textbf{Gen.} \\
\midrule
Monolithic             & 0.39 & 0.37 & 0.41 & 0.42 & 0.48 & 0.47 & 0.51 & 0.52 \\
SkillCom-Struct+Dedup  & \textbf{0.51} & \textbf{0.50} & \textbf{0.56} & \textbf{0.52} & \textbf{0.65} & \textbf{0.67} & \textbf{0.68} & \textbf{0.67} \\
SkillCom-Enrich        & 0.42 & 0.42 & 0.45 & 0.43 & 0.55 & 0.56 & 0.57 & 0.53 \\
\midrule
\multicolumn{9}{@{}l}{\scriptsize Tight$(2,24,150)$\; Med.$(3,36,225)$\; Def.$(4,48,300)$\; Gen.$(6,72,450)$} \\
\bottomrule
\end{tabular}
\end{table}

\begin{table}[t]
\centering
\caption{Budget sensitivity on MultiWOZ DST ($\mathrm{SNR}{=}7\,\text{dB}$).}
\label{tab:budget_dst}
\scriptsize
\setlength{\tabcolsep}{2.0pt}
\begin{tabular}{@{}l cccc cccc@{}}
\toprule
 & \multicolumn{4}{c}{\textbf{JGA}$\uparrow$} & \multicolumn{4}{c}{\textbf{Slot F1}$\uparrow$} \\
\cmidrule(lr){2-5} \cmidrule(lr){6-9}
\textbf{Method} & \scriptsize\textbf{Tight} & \scriptsize\textbf{Med.} & \scriptsize\textbf{Def.} & \scriptsize\textbf{Gen.} & \scriptsize\textbf{Tight} & \scriptsize\textbf{Med.} & \scriptsize\textbf{Def.} & \scriptsize\textbf{Gen.} \\
\midrule
Monolithic             & 0.05 & 0.03 & 0.04 & 0.04 & 0.09 & 0.04 & 0.05 & 0.05 \\
SkillCom-Enrich        & \textbf{0.08} & \textbf{0.08} & \textbf{0.08} & \textbf{0.09} & \textbf{0.33} & \textbf{0.36} & \textbf{0.42} & \textbf{0.47} \\
SkillCom-Struct+Dedup  & 0.01 & 0.03 & 0.03 & 0.03 & 0.14 & 0.23 & 0.27 & 0.27 \\
\midrule
\multicolumn{9}{@{}l}{\scriptsize Tight$(2,28,175)$\; Med.$(3,42,260)$\; Def.$(5,56,350)$\; Gen.$(7,84,525)$} \\
\bottomrule
\end{tabular}
\end{table}

SkillCom maintains a clear advantage at every budget level: on HotpotQA, SkillCom-Struct+Dedup attains F1~$=0.65$ even under the tightest budget $(B_u{=}2)$, compared with $0.48$ for the monolithic baseline.
Moreover, SkillCom converts additional budget into task performance more effectively; on MultiWOZ, the Slot~F1 of SkillCom-Enrich grows from $0.33$ to $0.47$ as the budget increases, while the monolithic baseline remains nearly flat ($0.04$--$0.09$), confirming that unit-level transmission makes each additional unit independently useful for downstream inference.

\section{Conclusion}
\label{sec:conclusion}

We presented SkillCom, a framework that decomposes LLM-based semantic communication into four independently replaceable skills connected through typed semantic-unit interfaces.
Experiments on HotpotQA and MultiWOZ show that SkillCom consistently outperforms a monolithic baseline in task accuracy, noise robustness, and budget efficiency, while skill ablation reveals task-dependent bottlenecks that are difficult to diagnose in monolithic systems.
These results suggest that explicit skill decomposition and unit-level transmission provide a more robust and diagnosable foundation for LLM-based semantic communication.


\end{document}